\documentclass[aps,amsmath]{revtex4}

\usepackage{graphicx}
\usepackage{dcolumn}
\usepackage{bm}

\begin{document}

\title{A Microcalorimeter and Bolometer Model}         

\author{M.~Galeazzi}
 \altaffiliation[Current Address: ]{University of Miami, 
Department of Physics, P.O. Box 248046, Coral Gables, FL 33124 (USA), 
e-mail: galeazzi@physics.miami.edu} 
\affiliation{University of Wisconsin, Physics Department, 1150 University Ave.,
Madison, WI 53706 USA //
and NASA/Goddard Space Flight Center, Greenbelt, MD 20771 U.S.A.}

\author{D.~McCammon}
\affiliation{University of Wisconsin, Physics Department, 1150 University Ave.,
Madison, WI 53706 USA}

\begin{abstract}

The standard non-equilibrium theory of noise in ideal bolometers and 
microcalorimeters fails to predict the performance of real devices due
to additional effects that become important at low temperature.
In this paper we extend the theory to include the most important of
these effects, and find that the performance of microcalorimeters operating
at 60~mK can be quantitatively predicted. We give a simple method for doing
the necessary calculations, borrowing the block diagram formalism from
electronic control theory.

\end{abstract}

\maketitle

\section*{Introduction}

A complete non-equilibrium theory for the noise in simple bolometers with
ideal resistive thermometers was given by J.~C.~Mather in
1982 \cite{mather82} and extended to microcalorimeter performance two 
years later \cite{moseley}.
Here we use the terms bolometer and calorimeter in the conventional sense, 
respectively indicating power detectors and integrating energy detectors.

This theory shows that the performance of these devices improves dramatically
as the operating temperature is reduced. However, at temperatures below
$\sim$200 mK, it becomes increasingly difficult to construct a bolometer
that behaves according to the ideal assumptions.
The resistance of the thermometer becomes dependent on readout power as
well as temperature and equilibration times between different parts of
the detector become significant. Thermodynamic fluctuations between
internal parts are then an additional noise source.
The physical description for most of these effects is straightforward,
but combining all of them into a detector model can be algebraically
daunting. 

Theoretical models that describe complex thermal architectures are 
necessary to understand the behavior of real devices, and
some groups have already extended the ``ideal'' model developed 
by J.~C.~Mather in 1982 to include some non-ideal effects in order to 
explain their experimental results \cite{hoevers,gildemeister,flei,piat}.
We developed a general bolometer and microcalorimeter model using the block 
diagram formalism of control theory. The formalism helps with the mechanics 
of the problem, while keeping the physical model reasonably transparent 
\cite{lee,nivelle,max_feed}.
In the model we have included the thermal decoupling between
the electron system and the phonon system in the thermometer, the so 
called {\em hot-electron} model, the thermal decoupling between the 
absorber and the thermometer, and non-ohmic behaviors of the thermometer
in addition to the hot-electron effect. 
The hot-electron model assumes that the resistance of the thermometer 
depends on the temperature of the electrons, and there is a thermal 
resistance between the electrons and the crystal lattice through which 
the bias power must flow, increasing the temperature of the electrons 
above the temperature of the lattice, and therefore changing the thermometer
resistance. This effect is well known in metals at low temperatures
and has recently been studied in semiconductors in the variable 
range-hopping regime \cite{wang,g_hot-el,dahai}.
The noise analysis incorporates terms for thermometer Johnson 
and $1/f$ noise, amplifier noise, load resistor Johnson noise, and 
thermodynamic fluctuations between the electron and phonon systems in 
the thermometer as well as between the absorber, the thermometer, and the 
heat sink.
In the model we also included the effect of thermometer non-ohmic
behavior, i.e. dependence of the thermometer resistance on the bias signal. 
This effect is particularly 
important when Transition Edge Sensors (TES) are used \cite{mark} 
and makes the model valuable for predicting the performance of this type 
of detector.

\section{The ideal model}

To help the reader understand the algebra of our model we decided to
start our analysis with an overview of the ideal model that has been
previously developed.
Despite their different applications, bolometers and 
microcalorimeters are very similar detectors and the theory of
their operation is largely the same.
The considerations of this paper apply to both
kinds of detectors unless otherwise specified and we will 
use the generic term ``detectors'' to refer to both.

\begin{figure}
\includegraphics{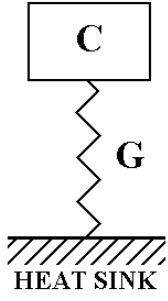}
\caption{\label{f_sketch1}
Thermal sketch of a bolometer or microcalorimeter}
\end{figure}

Typically a bolometer or a microcalorimeter is composed of
three parts: an absorber that converts the incident power or
energy into a temperature variation, a sensor that reads out
the temperature variation, and a thermal link between the detector
and a heat sink. 
The sensor is typically a resistor whose resistance strongly
depends on the temperature around the working point.
An ideal detector can be represented 
by a discrete absorber  of heat capacity $C$ in contact with the 
heat sink through a thermal conductivity $G$ (see Fig.~\ref{f_sketch1}),
and a thermometer always at the temperature of the absorber.
The thermometer sensitivity is specified by:
\begin{equation}
\alpha \equiv \frac{T}{R} \frac{dR}{dT},
\label{e_alpha}
\end{equation}
where $T$ is the detector temperature and $R$ is the sensor
resistance.
The thermal conductivity $G$ is defined as:
\begin{equation}
G\equiv \frac{dP}{dT},
\label{e_model1}
\end{equation}
where $P$ is the power dissipated into the detector. The conductivity 
$G$ can generally be expressed as a power law of the detector temperature 
$T$, i.e. $G=G_0 \cdot T ^{\beta}$.
Notice that numerically $G_0$ is equal to the thermal conductivity at
1~K, but dimensionally $G_0$ is a thermal conductivity divided by
a temperature to the $\beta$.

In equilibrium, with no other input power than the Joule power
$P$ used to read out the thermometer resistance, the equilibrium temperature 
of the detector $T$ is determined by integrating Eq.~\ref{e_model1} between 
the heat sink temperature $T_S$ and the detector temperature:
\begin{equation}
\int _{T_S}^{T} G(T') dT' =P(T).
\label{e_model1b}
\end{equation}

Assuming the power law expression for $G$ introduced before and integrating
it becomes:  
\begin{equation}
(T ^{\beta +1} -T_S ^{\beta +1})=\frac{\beta +1}{G_0}P(T).
\label{e_model2}
\end{equation}
It is important to remember, when calculating the equilibrium temperature,
that the power $P$ depends on the value of the sensor resistance and,
as a consequence, it depends on the temperature $T$, as explicitly 
indicated in Eq.~\ref{e_model2}.
To calculate the equilibrium temperature it is therefore necessary to solve the
system of equations represented by Eq.~\ref{e_model2}, the $P$~vs.~$R$ curve
and the $R$~vs.~$T$ curve. In general,the system must be solved numerically. 

Of interest from the point of view of the detector operation, is how 
the temperature rise $\Delta T$ above the equilibrium 
temperature depends on an external incident power $W$.
The power input to the detector ($W+P$) is partly
stored into the heat capacity of the detector and partly flows
to the heat sink through the thermal conductivity.
The equation that determines the generic temperature $T_D$ of the detector is
therefore:
\begin{equation}
C\frac{dT_D}{dt}+\int _{T_S}^{T_D} G(T') dT' =W+P(T_D),
\label{e_model2b}
\end{equation}
where we explicitly indicated that the bias power can be a function of the
temperature $T_D$ and where the quantities $T_D$, $W$, and $P$ can be a function
of time $t$.
We can express the generic detector temperature $T_D$ as a function of the equilibrium 
temperature $T$ defined  in Eq.~\ref{e_model2} as $T_D=T +\Delta T$. 
Eq.~\ref{e_model2b} then becomes:
\begin{equation}
C\frac{d(T +\Delta T)}{dt}+\int _{T_S}^{T} G(T') dT' 
+\int _{T}^{T +\Delta T} G(T')dT'=W+P(T +\Delta T).
\end{equation}
If we stay in the so called {\em small signal limit}, i.e.
we assume that $\Delta T$ is small compared to $T$,
we can expand the second integral to lowest order in $\Delta T/T$,
obtaining:
\begin{equation}
C\frac{d(T +\Delta T)}{dt}+\int _{T_S}^{T} G(T') dT' +G(T )\cdot \Delta T
=W+P(T )+\Delta P
\label{e_model2c}
\end{equation}
with $\Delta P=P(T +\Delta T)-P(T )$. Subtracting Eq.~\ref{e_model1b} 
from Eq.~\ref{e_model2c}, and considering that the equilibrium temperature
$T$ does not change with time, we obtain:
\begin{equation}
C\frac{d(\Delta T)}{dt}+G\cdot \Delta T=W+\Delta P,
\label{e_model3}
\end{equation}
where for simplicity we expressed $G\equiv G(T)$. 

\begin{figure}
\includegraphics{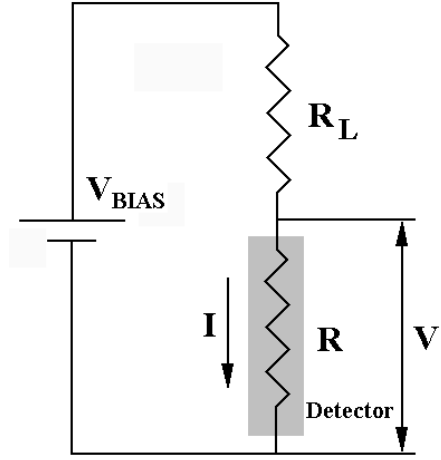}
\caption{\label{f_model1}
Typical detector readout circuit.}
\end{figure}

In general, the bias power will change with temperature, since $R$ changes,
and its expression depends on the bias source impedance.
A typical bias circuit is illustrated in Fig.~\ref{f_model1} where $R$ is the
thermometer resistance and $R_L$ is a load resistor. The most
commonly used bias conditions are near current bias ($R_L \gg R$) and near
voltage bias ($R_L \ll R$). More complex bias circuits are also used, and 
can always be represented by the circuit  of Fig.~\ref{f_model1} using
Thevenin equivalence theorems.
Differentiating the expression for the Joule power $P=I^2 R=V^2 /R$ 
and using the bias circuit of Fig.~\ref{f_model1} we obtain:
\begin{equation}
\Delta P=-\frac{P}{T}\frac{R-R_L}{R_L +R} \alpha \Delta T.
\label{e_model3a1}
\end{equation}

This term is generally referred to as the {\em electro-thermal
feedback} term and it often plays an important role in the response
of a detector. 
For simplicity in the small signal analytical calculations, we write
the electro-thermal feedback term as:
\begin{equation}
\Delta P=-G_{\mathit{ETF}} \Delta T,
\label{e_model3a3}
\end{equation}
where
\begin{equation}
G_{\mathit{ETF}}\equiv \frac{P}{T}\frac{R-R_L}{R_L +R} \alpha,
\end{equation}
so that Eq.~\ref{e_model3} becomes:
\begin{equation}
C\frac{d(\Delta T)}{dt}+(G+G_{\mathit{ETF}} )\cdot \Delta T=W
\label{e_model3b}
\end{equation}
or, introducing an equivalent thermal conductivity $G_{\mathit{eff}}=G+G_{\mathit{ETF}}$
(which we refer to as {\em effective thermal conductivity}):
\begin{equation}
C\frac{d(\Delta T)}{dt}+G_{\mathit{eff}} \cdot \Delta T=W.
\label{e_model4}
\end{equation}

The easiest way to solve this differential equation is using Fourier
transforms. The procedure is to use Fourier transforms to convert the
terms of Eq.~\ref{e_model4} to the frequency domain, solve the equation
in the frequency domain where it becomes a linear equation, then Fourier 
invert transform the result to the time domain. 
The advantage of solving Eq.~\ref{e_model4} in the frequency domain comes
from the fact that the expression $d\Delta T(t)/dt$ in the frequency
domain becomes $j\omega \Delta T(\omega )$, where we used the engineering
notation $j=\sqrt{-1}$. Equation~\ref{e_model4} in the frequency domain 
then becomes:
\begin{equation}
j\omega C\Delta T(\omega )+G_{\mathit{eff}} \cdot \Delta T(\omega )=W(\omega ).
\end{equation} 
whose solution is:
\begin{equation}
\Delta T(\omega)=\frac{1}{G_{\mathit{eff}}}\frac{1}{1+j\omega 
\tau _{\mathit{eff}}} W(\omega )
\label{e_model5}
\end{equation}
with $\tau _{\mathit{eff}}=C/G_{\mathit{eff}}$.

\begin{figure}
\includegraphics{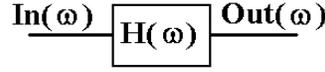}
\caption{\label{f_model2}
Block diagram representation of a system with transfer
function $H(\omega )$.}
\end{figure}

The detector system behaves as a low pass system, with 
time constant $\tau _{\mathit{eff}}$. For negative electrothermal feedback 
$G_{\mathit{ETF}}$ must be positive 
and the detector time constant is shortened. For positive feedback
$G_{\mathit{ETF}}$ is negative and the detector time constant is 
lengthened and, in the case
of $|G_{\mathit{ETF}}|$ bigger than $G$, the detector becomes unstable.
The sign of $G_{\mathit{ETF}}$ depends on the sign of $\alpha$ and on the
bias condition used (i.e., the ratio $R/R_L$).
In the small-signal (linear) limit considered here and in absence of 
amplifier noise, the signal has no effect on the detector performance.
However positive feedback reduces the effect of amplifier noise, while
negative feedback helps linearize the large signal gain, and improves
microcalorimeter resolution for large signals at high count rate. 
Since it can usually be arranged that amplifier noise is negligible,
these practical considerations normally favor negative feedback.
People then use current bias ($R<R_L$) for detectors with negative 
$\alpha$ and voltage bias ($R>R_L$) for detectors with positive $\alpha$.

In operating a detector, what is really detected is not directly
the temperature variation $\Delta T$, but the resistance variation
$\Delta R$, which is read out either as a voltage or current variation,
that is:
\begin{equation}
\Delta V=V\frac{\alpha}{T}\frac{R_L}{R_L +R} \Delta T
\label{e_atr1}
\end{equation}
\begin{equation}
\Delta I=-I\frac{\alpha}{T}\frac{R}{R_L +R} \Delta T.
\label{e_atr2}
\end{equation}
We can generically indicate the output signal as $X$ and
the relation between the output and the temperature as:
\begin{equation}
\frac{\Delta X}{X}=\alpha \cdot A_{tr} \frac{\Delta T}{T}
\label{e_atr3}
\end{equation}
where $A_{tr}$ is a dimensionless parameter that quantifies how much 
the output signal is sensitive to resistance changes and that we call
the transducer sensitivity. Numerically $A_{tr}$ is defined as:
\begin{equation}
A_{tr}\equiv \frac{R}{X}\frac{dX}{dR}.
\label{e_atr4}
\end{equation}
Notice that the expression of $A_{tr}$ can be easily derived from 
Eqs.~\ref{e_atr3}~and~\ref{e_atr1}~or~\ref{e_atr2} for voltage and current 
readout, and is always smaller or equal to unity for passive bias circuit
($R_L >0$).

The response of a detector is usually quantified by the 
{\em responsivity} $S(\omega )$, defined as:
\begin{equation}
S(\omega )=\frac{\Delta X(\omega )}{W(\omega )}
\end{equation}
that is, the responsivity characterizes the response of the detector, 
$\Delta X$, to an input power $W$.
In the ideal model just described we can combine 
Eqs.~\ref{e_model5}~and~\ref{e_atr3} to obtain:
\begin{equation}
\Delta X(\omega ) = \frac{1}{G_{\mathit{eff}}}\frac{1}{1+j\omega \tau _{\mathit{eff}}}
\frac{X \cdot \alpha \cdot A_{tr}}{T} \cdot W(\omega ),
\label{e_dx}
\end{equation}
and the responsivity is then equal to:
\begin{equation}
S(\omega)=\frac{1}{G_{\mathit{eff}}}\frac{1}{1+j\omega \tau _{\mathit{eff}}}
\frac{X \cdot \alpha \cdot A_{tr}}{T}.
\end{equation}

A detector at the working point is also often described by the 
complex dynamic impedance $Z(\omega )=dV(\omega )/dI(\omega )$. 
The dynamic impedance $Z(\omega )$ differs from the detector resistance
$R=V/I$ due to effect of the electro-thermal feedback. When
the current changes, the power dissipated into the detector changes too, 
therefore the temperature and the detector resistance change.
It is often useful to express the detector performance and characteristics
in terms of the dynamic impedance since it can be easily measured 
experimentally.

\begin{figure}
\includegraphics{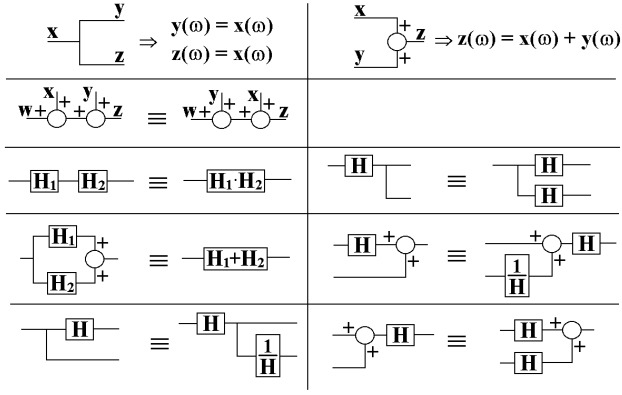}
\caption{\label{f_model3}
Some general operations with the block diagram algebra.}
\end{figure}

The calculation of the analytical expression of the dynamic impedance is
simple. Differentiating Ohm's law, $V=I\cdot R$, we obtain: 
\begin{equation}
dV=I\cdot dR+R\cdot dI. 
\label{e_z1}
\end{equation}
Using Eq.~\ref{e_model3} in the frequency domain
with $W=0$ and the definition of the thermometer sensitivity $\alpha$ in 
Eq.~\ref{e_alpha}, we obtain: 
\begin{equation}
dR=\frac{R}{T}\alpha dT=\frac{R\alpha}{GT} \frac{1}{1+j\omega \tau}dP.
\label{e_z2}
\end{equation}
with $\tau =C/G$. This $\tau$ is called the ``intrinsic'' or
``thermal'' time constant of the detector.
Differentiating the expression of the Joule power dissipated into the 
thermometer $P=V\cdot I$, we obtain:
\begin{equation}
dP=V\cdot dI+I\cdot dV,
\label{e_z2b}
\end{equation}
which, combined with Eqs.~\ref{e_z1}~and~\ref{e_z2} gives:
\begin{equation}
dV=R\cdot dI+I\frac{R\alpha}{GT}\frac{1}{1+j\omega \tau}(I\cdot dV+V\cdot dI).
\label{e_z3}
\end{equation}
Notice that in Eqs.~\ref{e_z1} through \ref{e_z3} most of the terms are 
function of the frequency $\omega$.
Solving Eq.~\ref{e_z3} we obtain:
\begin{equation}
Z(\omega )=\frac{dV(\omega )}{dI(\omega )}=R\frac{1+\frac{P\alpha}{GT}+j\omega \tau}
{1-\frac{P\alpha}{GT}+j\omega \tau}
=Z_0 \frac{1+j\omega \tau \frac{Z_0 +R}{2Z_0}}{1+j\omega \tau \frac{Z_0 +R}{2R}},
\label{e_z4}
\end{equation}
where we used the expression
\begin{equation}
Z_0 =Z(\omega =0)=R\frac{1+\frac{P\alpha}{GT}}{1-\frac{P\alpha}{GT}}.
\end{equation}
Notice that when $\omega \rightarrow \infty$, $Z \rightarrow R$.

The dynamic impedance, $Z(\omega )=dV/dI$, is easily measured 
experimentally. It can be determined most readily by adding a small 
A.C. signal to the bias voltage and measuring the {\em transfer 
function} of the detector $TF(\omega )$.  This is the ratio of amplitudes and 
relative phase between changes in the detector voltage and changes in 
the bias voltage as a function of frequency.  Most spectrum analyzers 
have the capability of measuring the complex ratio between two 
signals as a function of frequency, and can do this simultaneously 
over the frequency range of interest using a band-limited white noise 
source.  Signal averaging allows very precise measurement to be made 
while remaining in the small-signal limit.  The dynamic impedance is 
easily derived from the transfer function using the value of the load 
resistance and making appropriate corrections for stray electrical 
capacitance or inductance in the circuit. For example, in the bias 
circuit of Fig.~\ref{f_model1}, without stray capacitance, the impedance
is equal to: $Z(\omega )=R_L\cdot TF(\omega ) /(1-TF(\omega ))$.

It is then possible to determine values for many of the 
important parameters of the detector by fitting the real and 
imaginary parts of the transfer function by adjusting the thermal and 
electrical parameters in the expressions given in this paper.  This 
is very valuable for diagnosing performance problems or improving the 
design of detectors.  Note that when the thermometer temperature 
coefficient $\alpha$ is positive, the impedance can become infinite. 
It is then more convenient to work with the inverse quantity 
$1/Z(\omega )=dI/dV$.

In the case of a detector whose signal is read out as a voltage change, 
where the responsivity $S$ is defined as $S(\omega )=dV(\omega )/dW(\omega )$, 
we can also write:
\begin{equation}
S(\omega )=\frac{1}{2I}\frac{(Z_0 /R)-1}{(Z_0 /R_L )+1}\frac{1}
{1+j\omega \tau _{\mathit{eff}}}.
\end{equation}

At this point we want to introduce a useful technique
for analyzing the response of a bolometer or a microcalorimeter:
block diagram algebra. This technique is generally used
in electrical engineering to analyze feedback systems and it
is very useful when extending the theory
of bolometers and microcalorimeters to more complicated realistic systems.
The algebra of block diagrams and the language of control theory have been
successfully used before in the analysis of microcalorimeters 
and bolometer \cite{lee,nivelle,max_feed}. 
The basic idea is that a system with 
transfer function in the frequency domain equal to $H(\omega)$ 
is represented by the diagram of Fig.~\ref{f_model2}.
If an input $In(\omega )$ is applied to the system the
output is $Out(\omega )=H(\omega )\cdot In(\omega )$.
Complicated systems can always be reduced to the system of Fig.~\ref{f_model2}
using the block diagram algebra.
Fig.~\ref{f_model3} shows some of the common operations that will be 
used in this paper.
The procedure to solve the response of a system using the block
diagram algebra is then the following:
\begin{itemize}
\item Write the differential equations that define the system response.
\item Convert the equations to the frequency domain and, for each
equation define the individual system response and the input to 
that system.
\item Layout the block diagram that describes all the equations
together.
\item Use the block diagram algebra to reduce the block diagram to the 
form of Fig.~\ref{f_model2} that represents the system response in
the frequency domain.
\end{itemize}

\begin{figure}
\includegraphics{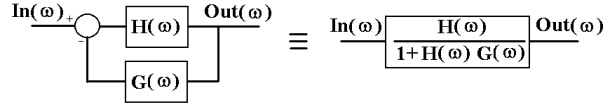}
\caption{\label{f_model4}
Block diagram representation of a feedback system.}
\end{figure}

This representation is particularly useful to deal with 
feedback systems, i.e. systems where the output is combined
to the input through a transfer function $G(\omega )$ as
in Fig.~\ref{f_model4}. In this case, whenever an external
input $In(\omega )$ is applied to the system the output
is:
\begin{equation}
Out(\omega )=\frac{H(\omega )}{1+H(\omega )G(\omega )} In(\omega )
=H_{\mathit{c-l}}(\omega ) 
\cdot In(\omega ),
\label{e_model6}
\end{equation}
where $H_{\mathit{c-l}}(\omega )$ is called the {\em closed loop transfer 
function}.

\begin{figure}
\includegraphics{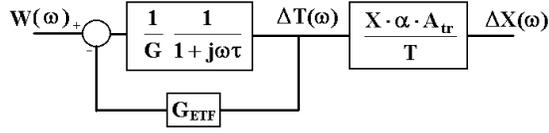}
\caption{\label{f_model5}
Block diagram representation of a detector.}
\end{figure}

Going back to the theory of bolometers and microcalorimeters,
we can write Eq.~\ref{e_model3b} as:
\begin{equation}
C\frac{d(\Delta T)}{dt}+G\cdot \Delta T=W-G_{\mathit{ETF}} \cdot \Delta T,
\label{e_bd}
\end{equation}
that, in the frequency domain, becomes:
\begin{equation}
j\omega C \Delta T+G\Delta T=W-G_{\mathit{ETF}} \Delta T.
\end{equation}
We now want to generate the block diagram describing this equation.
The left part represents the response of the system that we are 
analyzing (the output $\Delta T$ as a function of an input power), while the
right part represents the input to that system. The fact that the
input depends on the output $\Delta T$ is a consequence of feedback.
The left part of the equation represents a low pass system
with transfer function 
\begin{equation}
H(\omega )=\frac{1}{G} \frac{1}{1+j\omega \tau}
\end{equation}
The input consists of an external input $W$ minus
the output itself modified by the transfer function $G_{\mathit{ETF}}$.
This is a typical feedback system represented by the 
block diagram of Fig.~\ref{f_model5}, where we also included the 
conversion of $\Delta T$ to $\Delta X$.
If we now solve the block diagram using the block diagram algebra and
Eq.~\ref{e_model6}, we obtain:
\begin{equation}
\Delta X(\omega)=\frac{1}{G+G_{\mathit{ETF}}}\frac{1}{1+j\omega \frac{C}{G+G_{\mathit{ETF}}}} %
\frac{X \cdot \alpha \cdot A_{tr}}{T} W(\omega )
=\frac{1}{G_{\mathit{eff}}}\frac{1}{1+j\omega \tau _{\mathit{eff}}} %
\frac{X \cdot \alpha \cdot A_{tr}}{T}W(\omega )
\end{equation}
that is the same expression of Eq.~\ref{e_dx}.

\section{The Hot-electron Model}

\begin{figure}
\includegraphics{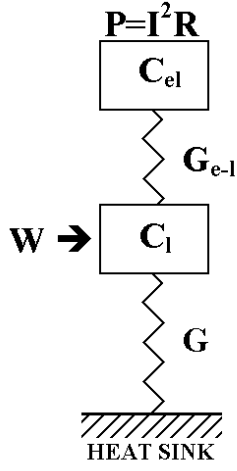}
\caption{\label{f_sketch2}
Thermal sketch of a bolometer or microcalorimeter in
the hot electron model}
\end{figure}

A first order correction to the standard theory of bolometers
and microcalorimeters is the introduction of the {\em hot-electron
model}. The model assumes that the thermal coupling between electrons 
and lattice in the sensor at low temperature is weaker than the
coupling between electrons, so that the electric power applied to
the electrons rises them to a higher temperature than the lattice.
This behavior is a known property of metals and has recently been
quantified in doped silicon \cite{dahai}, so that it 
affects both TES and semiconductor sensors.
The detector can therefore be described as composed of two 
different systems: the electron system and the
phonon or lattice system and the two are thermally connected by a 
thermal conductivity $G_{\mathit{e-l}}$.
We assume for models derived in this paper that the detector 
resistance responds to the temperature of its electron system, and 
that the Joule power of the bias is dissipated there.  For economy of 
presentation, the models derived here assume the input power enters 
through the absorber phonon system, which is then thermally connected 
to the thermometer lattice, and further to the heat sink through the 
thermal conductivity $G$ (see Fig.~\ref{f_sketch2}). 
There are important classes of 
detectors where signal power is absorbed directly in the electron 
system of the thermometer or absorber, and the primary thermal path 
to the heat sink could be from the absorber lattice or either 
electron system.  In the general case, these all result in different 
thermal circuits, and the block diagrams must be modified 
accordingly.  In the approximation of this section, the phonon system 
includes both the absorber and the phonons in the thermometer (we 
will discuss later the case of a decoupled absorber).

In equilibrium with no external power, the electron system is
at a higher temperature than the phonon system is and the
Joule power flows from the electron system to the phonon 
system and from there to the heat sink.
The equilibrium temperature of the two systems without
any signal power applied can be calculated in a
way similar to that used for the simple model described 
in the previous paragraph.
As reported in the literature, the thermal conductivity
between electrons and phonons can be described as a power law 
of the electron temperature $T_e$ \cite{wang,g_hot-el}:
\begin{equation}
G_{\mathit{e-l}}=G_{0e} \cdot T_e ^{\beta _e}.
\label{e_hot0}
\end{equation}

From the definition of thermal conductivity, we also have:
\begin{equation}
G_{\mathit{e-l}}=\frac{dP}{dT_e}
\end{equation}
and if we combine the two and integrate from the lattice
temperature $T_l$ to the equilibrium electron temperature $T_e$
we obtain
\begin{equation}
T_e ^{\beta _e+1}=\frac{\beta _e+1}{G_{0e}} \cdot P(T_e ) +T_l ^{\beta _e+1}
\label{e_hot1}
\end{equation}
where we explicitly indicated the dependence of the power $P$ 
on the electron temperature $T_e$.
The equilibrium temperature of the lattice system is still
determined by Eq.~\ref{e_model2} that, in this case, can be
written as:
\begin{equation}
(T_l ^{\beta +1} -T_S ^{\beta +1})=\frac{\beta +1}{G_0}P(T_e).
\label{e_hot1b}
\end{equation}
Equations~\ref{e_hot1} and \ref{e_hot1b} represent a system with 
two variables $T_e$ and $T_l$ that can be solved numerically.

Here we are considering detectors where the external power is 
absorbed in the phonon system, and the sensitivity of the detector 
can be strongly affected by the reduced sensitivity of $T_e$ to changes 
in $T_l$ introduced by the equilibrium difference of these temperatures 
and non-linear nature of $G_{\mathit{e-l}}$. We consider these effects in two 
steps.  A first approximation is to assume that the heat capacity of 
the electron system is negligible.  This case can be solved easily, 
and it is sufficient in many cases.  We will then derive the general 
result for $C_e \ne 0$.

\subsection{Hot Electron Model with $C_e$ = 0}

If the electron system heat capacity $C_e$ can be neglected, 
the dependence of the electron 
temperature on the lattice temperature is simply
determined by Eq.~\ref{e_hot1}.
When the temperature of the lattice system 
changes by ${\Delta T_l}$ the temperature of the
electron system will instantly change by $\Delta T_e$ 
and Eq.~\ref{e_hot1} becomes:
\begin{equation}
(T_e +\Delta T_e )^{\beta _e+1}=\frac{\beta _e+1}{G_{0e}} \cdot P(T_e +\Delta T_e ) 
+(T_l +\Delta T_l )^{\beta _e+1}.
\label{e_hot2}
\end{equation}
If we subtract Eq.~\ref{e_hot1} from Eq.~\ref{e_hot2} we obtain
\begin{equation}
(T_e +\Delta T_e )^{\beta _e+1} - T_e ^{\beta _e+1}=\frac{\beta _e+1}{G_{0e}} 
(P(T_e +\Delta T_e )-P(T_e )) +(T_l +\Delta T_l )^{\beta _e+1} -T_l ^{\beta _e+1}.
\label{e_hot3}
\end{equation}

Assuming that $\Delta T_e \ll T_e$ and $\Delta T_l \ll T_l$
we can expand Eq.~\ref{e_hot3} to lowest order in $\Delta T_e / T_e$ 
and $\Delta T_l / T_l$, obtaining:
\begin{equation}
(1 +(\beta _e+1) \frac{\Delta T_e}{T_e} ) \cdot T_e ^{\beta _e+1}
-T_e ^{\beta _e+1} =\frac{\beta _e+1}{G_{0e}} \cdot \Delta P 
+(1 +(\beta _e+1) \frac{\Delta T_l}{T_l} ) \cdot T_l ^{\beta _e+1}
-T_l ^{\beta _e+1}
\end{equation}
that reduces to:
\begin{equation}
T_e ^{\beta _e} \cdot \Delta T_e = \frac{\Delta P}{G_{0e}} 
+T_l ^{\beta _e} \cdot \Delta T_l.
\label{e_hot4}
\end{equation}

We already calculated the change in Joule power $\Delta P$ in 
Eqs.~\ref{e_model3a1}~and~\ref{e_model3a3}, that in the hot electron case
depends on the change in electron temperature $\Delta T_e$:
\begin{equation}
\Delta P=-G_{\mathit{ETF}} \Delta T_e,
\label{e_hot5}
\end{equation}
therefore
\begin{equation}
\Delta T_e = \frac{T_l ^{\beta _e}}{T_e ^{\beta _e}+\frac{G_{\mathit{ETF}}}{G_{0e}}} 
\Delta T_l =
\frac{G_{\mathit{e-l}}(T_l )}{G_{\mathit{e-l}}(T_e )+G_{\mathit{ETF}}}\Delta T_l
 =\frac{G_{\mathit{e-l}}(T_l )}{G_{\mathit{e-l}}(T_e )}\frac{1}{1+
\frac{G_{\mathit{ETF}}}{G_{\mathit{e-l}}(T_e )}}\Delta T_l=A_{\mathit{e-l}} \Delta T_l
\label{e_hot6}
\end{equation}
where $G_{\mathit{e-l}}(T_e )$ is the electron-lattice thermal conductivity 
calculated at the electron temperature and $G_{\mathit{e-l}}(T_l )$ is the
electron-lattice thermal conductivity calculated at the lattice temperature,
and:
\begin{equation}
A_{\mathit{e-l}} =\frac{\Delta T_e}{\Delta T_l}=
\frac{G_{\mathit{e-l}}(T_l )}{G_{\mathit{e-l}}(T_e )}\frac{1}{1+
\frac{G_{\mathit{ETF}}}{G_{\mathit{e-l}}(T_e )}}.
\label{e_hot7}
\end{equation}
The quantity  $A_{\mathit{e-l}}$ is adimensional and represents the temperature 
sensitivity of the thermometer. When $A_{\mathit{e-l}}=1$, the thermometer is 
completely sensitive to temperature changes in the lattice system, when
$A_{\mathit{e-l}}=0$ the thermometer is completely insensitive to temperature
changes in the lattice.

\begin{figure}
\includegraphics{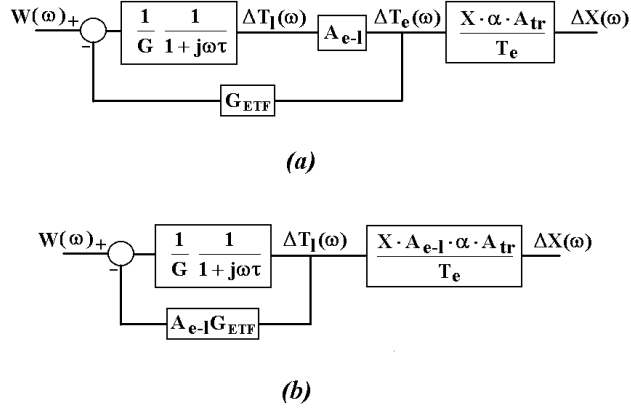}
\caption{\label{f_hot1}
Block diagram representation of a detector
using the hot electron model with $C_e =0$. (a) Block diagram 
as derived from the the equations that describe the detector.
Notice that the representation of the $ETF$ as acting on the 
lattice system of the sensor is due to the fact that we are 
assuming that $C_e =0$. In general, if $C_e \ne 0$, the $ETF$
is an electric effect and acts on the electron system.
(b) Equivalent representation to highlight the effect of the 
term $A_{e-l}$.}
\end{figure}

We now want to represent the detector using the block diagram algebra.
The detector behavior is described by Eqs.~\ref{e_atr3},~\ref{e_bd}, and~\ref{e_hot6},
that in the frequency domain can be written as:
\begin{equation}
j\omega C\Delta T_l+G\Delta T_l=W-G_{\mathit{ETF}} \Delta T_e,
\end{equation}
\begin{equation}
\Delta T_e=A_{e-l}\Delta T_l,
\end{equation}
and
\begin{equation}
\Delta X=\frac{X\alpha A_{tr}}{T_e} \Delta T_e.
\end{equation}
Converting these three equations in block diagram algebra and connecting
the blocks of the algebra together we obtain the representation of 
Fig.~\ref{f_hot1}a. 
With some simple algebra, the diagram 
is equivalent to that of Fig.~\ref{f_hot1}b, and considering 
that $G_{\mathit{ETF}} \propto \alpha$, the hot electron model with negligible 
heat capacity of the electron system is then equivalent to the 
standard model with the substitutions:
\begin{equation}
\alpha \rightarrow \alpha _{\mathit{eff}}=A_{\mathit{e-l}} \cdot \alpha
\end{equation}
\begin{equation}
T\rightarrow T_e
\end{equation}
and therefore the responsivity of the detector becomes
\begin{equation}
S(\omega )=\frac{1}{(G+A_{\mathit{e-l}}G_{\mathit{ETF}})}
\frac{1}{(1+j\omega \tau _{\mathit{eff}})}
\frac{A_{\mathit{e-l}} \cdot \alpha \cdot X \cdot A_{tr}}{T_e}.
\label{e_hot8}
\end{equation}
with $\tau_{\mathit{eff}}=C_l /(G+A_{\mathit{e-l}}G_{\mathit{ETF}})$, where $C_l$ is the lattice 
heat capacity.

\begin{figure}
\includegraphics{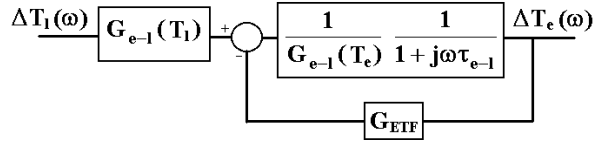}
\caption{\label{f_hotb1}
Representation of the electron system in the
hot electron model with $C_e \ne 0$.}
\end{figure}

\subsection{Hot Electron Model with C$_e$ $\ne$ 0}

If the heat capacity of the electron system is not negligible,
the electron temperature is defined (in analogy to Eq.~\ref{e_model2b}) 
by:
\begin{equation}
C_e\frac{dT_e}{dt}+\int _{T_l}^{T_e} G_{\mathit{e-l}} (T')dT'=P(T_e )
\label{e_hotb1}
\end{equation}
with $G_{\mathit{e-l}}$ defined by Eq.~\ref{e_hot0}.
What we are interested in is the rise of the electron temperature
$\Delta T_e$ above equilibrium when the 
lattice temperature rises by $\Delta T_l$.
Eq.~\ref{e_hotb1} then becomes
\begin{equation}
C_e\frac{d(T_e +\Delta T_e)}{dt}+
\int _{T_l +\Delta T_l}^{T_e +\Delta T_e} G_{\mathit{e-l}} (T')dT'
=P(T_e +\Delta T_e ).
\label{e_hotb2}
\end{equation}
Subtracting Eq.~\ref{e_hotb1} from Eq.~\ref{e_hotb2} we obtain
\begin{equation}
C_e\frac{d(\Delta T_e)}{dt}+\int _{T_e}^{T_e +\Delta T_e} G_{\mathit{e-l}} (T')dT'
-\int _{T_l}^{T_l +\Delta T_l} G_{\mathit{e-l}} (T')dT'=\Delta P
\label{e_hotb3}
\end{equation}
which, using Eq.~\ref{e_hot0} and expanding the result
to lowest order in $\Delta T_e /T_e$ and $\Delta T_l /T_l$,
becomes
\begin{equation}
C_e\frac{d(\Delta T_e)}{dt}+G_{\mathit{e-l}} (T_e )\Delta T_e=
G_{\mathit{e-l}} (T_l ) \Delta T_l -G_{\mathit{ETF}} \Delta T_e ,
\label{e_hotb4}
\end{equation}
with $G_{\mathit{ETF}}$ defined by Eq.~\ref{e_hot5}.
The system of Eq.~\ref{e_hotb4} is represented by the block diagram of 
Fig.~\ref{f_hotb1} with $\tau _{\mathit{e-l}} =C_e /G_{\mathit{e-l}}(T_e )$, which has the 
solution
\begin{equation}
\Delta T_e (\omega)=A_{\mathit{e-l}}\frac{1}{1+j\omega \tau _e}\Delta T_l ,
\label{e_hotb5}
\end{equation}
with $\tau _e =C_e /(G_{\mathit{e-l}}(T_e )+G_{\mathit{ETF}} )$ and $A_{\mathit{e-l}}$ defined by
Eq.~\ref{e_hot7}. Notice that Eq.~\ref{e_hotb5}
reduces to Eq.~\ref{e_hot6} if $C_e =0$.

\begin{figure}
\includegraphics{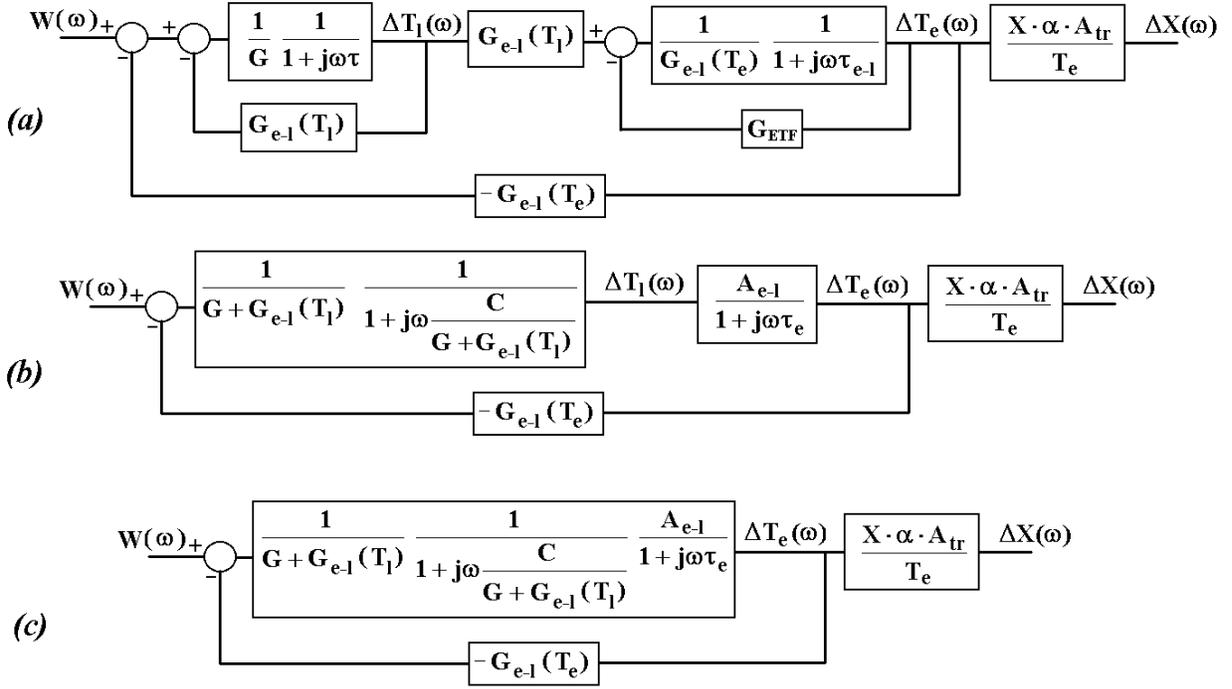}
\caption{\label{f_hotb2}
Block diagram representation of a detector
using the hot electron model with $C_e \ne 0$. (a) Block diagram
as derived from the equations describing the system. (b),(c)
Intermediate steps for the solution of the block diagram 
representation.
}
\end{figure}

The behavior of the lattice system is still regulated by Eqs.~\ref{e_model2b},
\ref{e_model2c} and \ref{e_model3}, with the substitutions of $C_l$ for $C$
and of $T_l$ for $T$, i.e.:
\begin{equation}
C_l \frac{d(\Delta T_l)}{dt}+G\cdot \Delta T_l=W+\Delta P_l.
\label{e_hotb6}
\end{equation}
The power $P_l$ is  the power flowing from the electron system to the lattice
system through the thermal conductivity $G_{\mathit{e-l}}$:
\begin{equation}
P_l =\int _{T_l}^{T_e} G_{\mathit{e-l}}(T')dT' .
\end{equation}
Therefore
\begin{equation}
\Delta P_l =\int _{T_e}^{T_e +\Delta T_e} G_{\mathit{e-l}}(T')dT'
-\int _{T_l}^{T_l +\Delta T_l} G_{\mathit{e-l}}(T')dT' ,
\end{equation}
which, considering the expression of Eq.~\ref{e_hot0} for
the thermal conductivity and expanding the result
to lowest order in $\Delta T_e /T_e$ and $\Delta T_l /T_l$, becomes:
\begin{equation}
\Delta P_l =G_{\mathit{e-l}} (T_e )\Delta T_e - G_{\mathit{e-l}} (T_l )\Delta T_l.
\end{equation}
Eq.~\ref{e_hotb6} then becomes:
\begin{equation}
C_l \frac{d(\Delta T_l)}{dt}+G\cdot \Delta T_l=W+G_{\mathit{e-l}} (T_e )\Delta T_e 
-G_{\mathit{e-l}} (T_l )\Delta T_l.
\label{e_hotb7}
\end{equation}
Equations~\ref{e_hotb4}~and~\ref{e_hotb7} can be written in the frequency
domain as:
\begin{equation}
j\omega C_e \Delta T_e +G_{\mathit{e-l}} (T_e )\Delta T_e=
G_{\mathit{e-l}} (T_l )\Delta T_l -G_{\mathit{ETF}} \Delta T_e
\label{new1}
\end{equation}
and
\begin{equation}
j\omega C_l \Delta T_l +G\cdot \Delta T_l=W+G_{\mathit{e-l}} (T_e )\Delta T_e 
-G_{\mathit{e-l}} (T_l )\Delta T_l,
\label{new2}
\end{equation}
and are represented by the block diagram of Fig.~\ref{f_hotb2}a.
The diagram can be solved to obtain an analytical expression
for the detector responsivity. 
In Figs.~\ref{f_hotb2}b~and~\ref{f_hotb2}c we show two
intermediate steps in the solution of the block diagram
algebra. The detector responsivity is then equal to:
\begin{equation}
S(\omega )=\frac{1}{G_{\mathit{ETF}} A_{\mathit{e-l}}(1+j\omega \frac{C_e}{G_{\mathit{ETF}}})+
G(1+j\omega \tau_l )(1+j\omega \tau _{e})}
\frac{A_{\mathit{e-l}} \cdot X \cdot \alpha \cdot A_{tr}}{T_e}.
\label{e_hotb8}
\end{equation}
with $\tau_l =C_l /G$. Notice that in the case of $C_e =0$ this 
expression reduces to Eq.~\ref{e_hot8}, i.e. the hot electron 
model with negligible electron heat capacity, as expected.

Moreover, in the case of
$G_{0e}\rightarrow \infty$, i.e. electrons and phonons can
be thermally considered as a single system the responsivity becomes:
\begin{equation}
S(\omega )=\frac{1}{(G+G_{\mathit{ETF}})}\frac{1}{\left( 1+j\omega \frac{C_l +C_e}{G+G_{\mathit{ETF}}}\right) } %
\frac{X \cdot A_{tr}\cdot \alpha}{T_{e}}.
\label{e_hotb9}
\end{equation}
This is just the ideal responsivity of a bolometer or
microcalorimeter with thermal conductivity $G$, temperature $T_e$ and 
heat capacity $C=C_l +C_e$.

\begin{figure}
\includegraphics{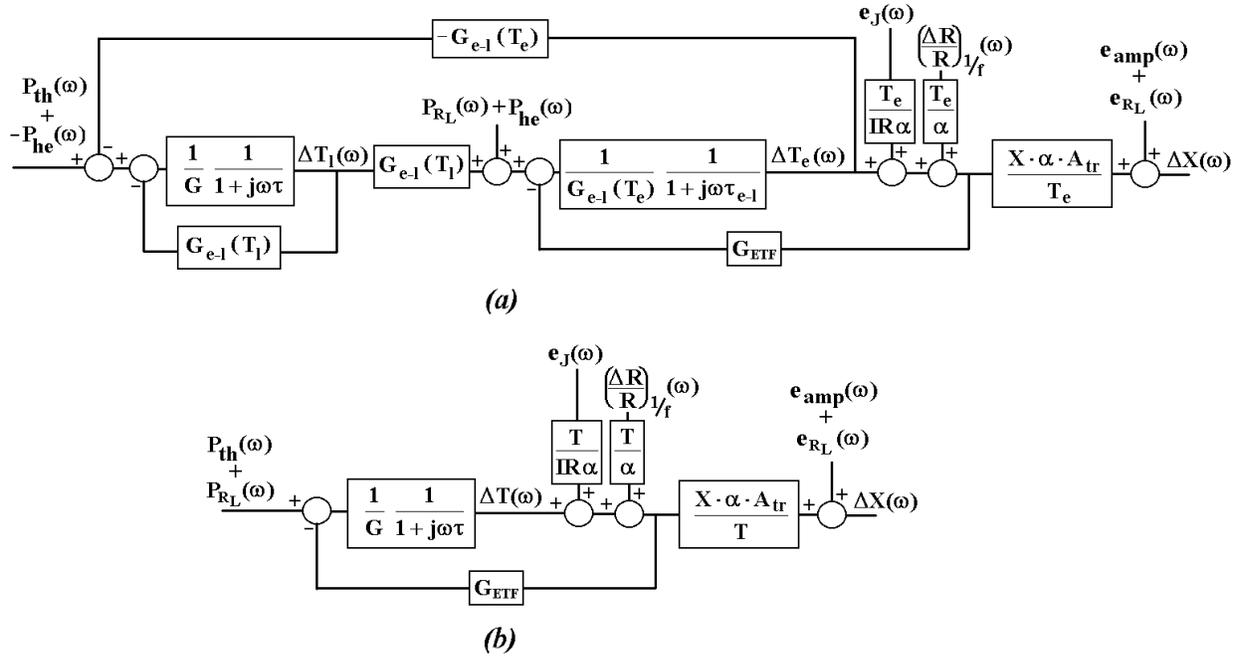}
\caption{\label{f_nep1}
Block diagram representation of the noise in a
a detector using the hot electron model(a) and equivalent representation
for the ideal model (b). Notice that if the output
$X$ is a current, the load resistor noise that adds to the output is
represented by $i_{R_L}$.}
\end{figure}

In analogy to Eqs.~\ref{e_z1} through \ref{e_z4}, we can
also calculate the dynamic impedance of the detector.
We can write Eqs.~\ref{new1} and \ref{new2} without external power $W$ and 
explicitly using the symbol $\Delta P$ for the change in Joule power:
Converting Eqs.~\ref{e_hotb4}~and~\ref{e_hotb7} to the frequency
domain we obtain:
\begin{equation}
j\omega C_e \Delta T_e +G_{\mathit{e-l}} (T_e )\Delta T_e=G_{\mathit{e-l}}(T_l ) %
\Delta T_l + \Delta P
\label{e_ze1}
\end{equation}
and
\begin{equation}
j\omega C_l \Delta T_l +G \Delta T_l=G_{\mathit{e-l}} (T_e )\Delta T_e %
- G_{\mathit{e-l}} (T_l )\Delta T_l.
\label{e_ze2}
\end{equation}

Combining 
Eqs.~\ref{e_alpha},~\ref{e_z1},~\ref{e_z2b},~\ref{e_ze1}~and~\ref{e_ze2},
we then obtain:
\begin{equation}
Z(\omega )=R\frac{ %
(G+G_{\mathit{e-l}}(T_l )+j\omega C_l) %
\left( G_{\mathit{e-l}}(T_e )+j\omega C_e +\frac{P\alpha}{T_e}\right) %
-G_{\mathit{e-l}}(T_e ) \cdot G_{\mathit{e-l}}(T_l )
}{
(G+G_{\mathit{e-l}}(T_l )+j\omega C_l) %
\left( G_{\mathit{e-l}}(T_e )+j\omega C_e -\frac{P\alpha}{T_e}\right) %
-G_{\mathit{e-l}}(T_e ) \cdot G_{\mathit{e-l}}(T_l )
}
\end{equation}

\section{Noise sources}

There are several noise sources that affect the performance
of bolometers and microcalorimeters, most of which have
already been taken into account by Mather in 1982 \cite{mather82}.
These include the Johnson noise of the sensor,
the thermal noise due to the thermal link between the detector
and the heat sink (also referred to as phonon noise), the Johnson 
noise of the load resistor
used in the bias circuit and the noise of the read-out electronics 
(amplifier noise). 
In his paper Mather also mentions a
$1/f$ noise contribution that seems to be more related to
the sensor characteristics. This noise has been 
studied and quantified for silicon implanted thermistors
by Han et al. in 1998 \cite{han}. 

The effect of the noise on the detector performance is
generally quantified by the {\em Noise Equivalent Power} ($NEP$).
The $NEP$ corresponds to the power $W(\omega )$ that would be
necessary as input of the detector to generate an output 
$\Delta X(\omega )$ equal to the output
generated by the noise. The $NEP$ is calculated as the ratio between
the output $\Delta X(\omega)$ generated by the noise 
and the responsivity of the detector $S(\omega )$.
In the case of bolometers, the $NEP$ directly quantifies
the limit of the bolometer in detecting a power signal at
frequency $\omega$. In the case of microcalorimeters
the $NEP$ is related to the best possible energy resolution
of the microcalorimeter by the expression \cite{moseley}:
\begin{equation}
\Delta E_{RMS}=\frac{1}{\sqrt{\int _0 ^{\infty} 
\frac{2 d\omega}{\pi NEP^2 (\omega )}}}.
\label{e_nep1}
\end{equation}

Here we want to analyze the effect of the noise on the detector performance
in the picture of the hot electron model.
The introduction of the hot electron model has two main effects:
it changes the $NEP$ of the noise sources and it introduces 
a new noise term, that is the thermal noise due to the thermal 
fluctuations between the lattice and electron systems.

\subsection{Effect of the Hot Electron Model on the Noise}

The different noise contributions affect the detector in
different ways. In particular, the thermal noise corresponds to 
a power noise on the lattice system. The Johnson noise is calculated as
a voltage fluctuation but can be introduced in the model as an electron
temperature noise term. The $1/f$ noise is calculated as a fluctuation 
in the value of the resistance 
but can be described as electron temperature noise term as well.
The load resistor noise can be described as a noise that adds to the 
output signal and also generates a Joule power noise on the electron 
system. The amplifier noise adds directly to the output signal.
In Fig.~\ref{f_nep1}a the contribution of the different
noise terms in the microcalorimeter are shown. 
The same noise sources in the
ideal model scenario are shown in Fig.~\ref{f_nep1}b \cite{max_feed}.
Dimensionally, the thermal noise $P_{th}$ is a power spectral
density (in units of $W/\sqrt{Hz}$), the
Johnson noise $e_J$ and the load resistor noise $e_{R_L}$
are voltage spectral densities ($V/\sqrt{Hz}$), the $1/f$ noise 
$\left(\frac{\Delta R}{R}\right) _{1/f}$ has dimensions 
$Hz^{-1/2}$ and the amplifier noise 
$e_{amp}$ has the dimension of the transducer output $X$ divided 
by square root of frequency ($V/\sqrt{Hz}$ or $A/\sqrt{Hz}$).
The thermal noise has been calculated quantitatively by
Mather in 1982 (assuming diffusive thermal conductivity) and is equal to:
\begin{equation}
P_{th}=\sqrt{4k_b G T_l ^2}\left( \frac{\int _{T_S}^{T_l} %
\frac{(T^{\prime}k(T^{\prime}))^2}{(T_l k(T_l ))^2} dT^{\prime}} %
{\int _{T_S}^{T_l} \frac{k(T^{\prime})}{k(T_l )}dT^{\prime}}\right) ^{1/2} 
\end{equation}
where $k_b $ is the Boltzmann constant and $k(T')$ is the 
function describing the temperature dependence of the 
thermal conductivity of the heat link material.

The Johnson noise of the sensor resistance is simply
described by:
\begin{equation}
e_J =\sqrt{4k_b T_e R}
\end{equation}

The Load resistor noise can be represented by a voltage noise across
the detector, equal to:
\begin{equation}
e_{R_L}=\sqrt{4k_b T_S R_L}\frac{R}{R_L +R}
\end{equation}
where we assumed that the electrical circuit is heat sunk
at the temperature $T_S$. This noise adds directly to 
the output signal as a voltage $e_{R_L}$ or as a current 
$i_{R_L}=e_{R_L}/R$, and generates Joule power noise in the electron
system $P_{R_L}=2I\cdot e_{R_L}$ (see Fig.~\ref{f_nep1}).

The $1/f$ noise is, by definition, frequency dependent, and it is usually
described as a fluctuation in the value of the resistance:
\begin{equation}
\left(\frac{\Delta R}{R}\right) _{1/f}\propto \frac{1}{\sqrt{\omega}}.
\end{equation}

Solving the block diagram of Fig.~\ref{f_nep1} independently
for each noise contribution and using the expression
of $S(\omega )$ of Eq.~\ref{e_hotb8} we obtain:
\begin{equation}
NEP_{th}=P_{th}
\label{e_nepth}
\end{equation}
\begin{equation}
NEP_J =\sqrt{\frac{4k_b T_e}{P\alpha ^2}}\frac{T_e ^{\beta _e +1}} %
{T_l ^{\beta _e}} \left( G(1+j\omega \tau _l )(1+j\omega \tau _{\mathit{e-l}}) %
+j\omega C_e \frac{T_l ^{\beta _e}} {T_e ^{\beta _e}} \right)
\label{e_nepj}
\end{equation}
\begin{equation}
NEP_{R_L} =\frac{e_{R_L}}{S(\omega )}+\frac{2Ie_{R_L}}{G_{\mathit{e-l}}(T_l )} %
\left( G+G_{\mathit{e-l}} (T_l ) + j\omega C_l \right)
\end{equation}
\begin{equation}
NEP_{1/f} =\left(\frac{\Delta R}{R}\right) _{1/f}\frac{T_e}{\alpha}
\frac{T_e ^{\beta _e}}{T_l ^{\beta _e}} %
\left( G(1+j\omega \tau _l ) (1+j\omega \tau _{\mathit{e-l}}) +j\omega C_e %
\frac{T_l ^{\beta _e}} {T_e ^{\beta _e}}\right)
\label{e_nep1f}
\end{equation}
\begin{equation}
NEP_{amp}=\frac{e_{amp}}{S(\omega)}.
\end{equation}

Notice that the $NEP$ due to the read-out electronics and to the
load resistor are the
only terms that depend on the electro-thermal feedback.
Therefore, if these terms are small compared to 
the other contributions, as it is usually the case, the electro-thermal 
feedback changes the time constant of the detector, but does not
affect $NEP(\omega )$.

The expression of the $NEP$ in the case of negligible electron 
heat capacity is easily derived using $C_e =0$. Notice that in 
the limit of $G_{0e}\rightarrow \infty$, Eqs.~\ref{e_nepth} and
\ref{e_nepj} reduce to the ideal expressions calculated 
by Mather \cite{mather82}:
\begin{equation}
NEP_{th}=P_{th}
\end{equation}
\begin{equation}
NEP_J =\frac{\sqrt{4k_bT_e}}{\sqrt{P}\alpha}T_e G
\left( 1+j\omega \frac{C_l +C_e}{G}\right) .
\end{equation}

\subsection{Thermal Noise Due to Hot Electron Decoupling}

The hot electron model also introduces an extra
noise term in addition to those just considered.
This is due to power fluctuations between the lattice and electron system.
The magnitude of these fluctuations depends in part on the physics of 
the electron-phonon decoupling.  
A simple expression appropriate for ``radiative'' energy transfer 
was calculated by Boyle and Rodger in 1959 \cite{boyle}:
\begin{equation}
P_{he}=\sqrt{2k_b G_{\mathit{e-l}}(T_e ) \frac{T_e ^5 +T_l ^5}{T_e ^3}}.
\end{equation}
A more rigorous expression for electron-phonon decoupling has 
also been calculated by Golwala et al. in 1997 \cite{golwala}.

\begin{figure}
\includegraphics{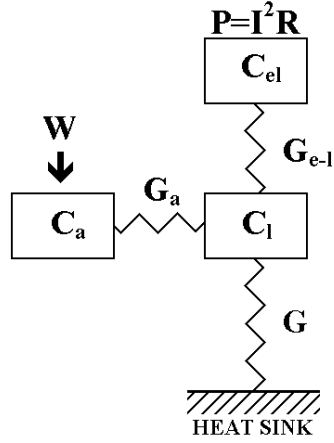}
\caption{\label{f_sketch3}
Thermal sketch of a bolometer or microcalorimeter
in the case of absorber decoupling and hot electron model}
\end{figure}

Notice that these fluctuations transport power from
the lattice system to the electron system and viceversa,
therefore if a power $P_{he}$ adds to the
electron system, the same power $P_{he}$ is subtracted
from the lattice system.
The effect is shown in the block diagram of Fig.~\ref{f_nep1}.
Solving the block diagram for the hot electron noise we obtain:
\begin{equation}
NEP_{he}=P_{he}\frac{G}{G_{\mathit{e-l}}(T_l )} %
(1+j\omega \tau _l),
\end{equation}
where $\tau _l$ has been previously defined as $\tau _l=C_l / G$.
This expression does not depend on $C_e$, therefore
is valid also for the case $C_e =0$.
Moreover, if $G_{0e}\rightarrow \infty$, this term
is zero, as expected.

\section{Absorber Decoupling}

Another aspect that may affect the performance of
bolometers and microcalorimeters that we want to
study is the effect of the absorber thermal conductivity.
Most of the detectors are built with absorber and sensor
as different entities connected by epoxy or other
material with a thermal conductivity $G_a$.
Depending on the experimental setup, there are 
different configurations that must be used to describe the
thermal system. For example, the thermal link to the heat
sink can be through the absorber or the thermometer and 
the absorber can be in thermal connection with the lattice 
system (when an electrical insulating material is used) or 
the electron system (when a conducting material is used). 

What we want to analyze here is the case in which 
the detector is connected to the heat sink through
the thermometer lattice system and the absorber is connected
to the lattice system of the thermometer. In this case, the 
external power hits the absorber, is released to the lattice 
system and then detected in the electron system 
(see Fig.~\ref{f_sketch3}).
We assume that the absorber has a heat capacity $C_a$.
Notice that the analytical tools that we give here can be
easily used to quantify the behavior of any other configuration.

\subsection{Responsivity and Dynamic Impedance}

\begin{figure}
\includegraphics{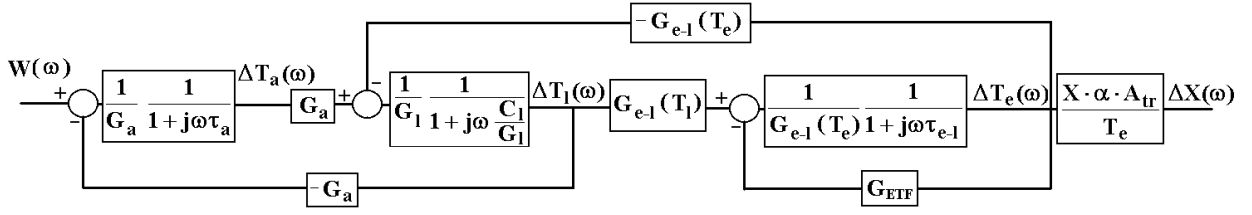}
\caption{\label{f_abs1}
Block diagram representation of a detector
with a finite thermal conductivity between absorber
and lattice system. We have used the notation 
$G_l=G+G_a +G_{\mathit{e-l}}(T_l )$. Notice that this implicitly
integrates the heat relief for the lattice system provided by the
electron and absorber decouling into the lattice response 
function. This is different from what was done before 
in Fig.~\ref{f_hotb2}, where the heat relief was explicitly 
reported in the block diagram as a feedback effect. The two
descriptions are equivalent. We used the implicit description
here to compact the block diagram algebra.}
\end{figure}

In equilibrium, with no other power input than the Joule
power in the sensor, there is no power flow through the
thermal link $G_a$ and therefore the temperature of the
absorber is equal to the lattice temperature $T_a =T_l$.
If an external power $W$ is applied to the absorber, the
detector is described in the frequency domain by the set
of equations:
\begin{equation}
j\omega C_a \Delta T_a+G_a \Delta T_a = W+G_a \Delta T_l
\label{e_ab1}
\end{equation}
\begin{equation}
j\omega C_l \Delta T_l+(G+G_a +G_{\mathit{e-l}}(T_l ))\Delta T_l = %
G_a \Delta T_a +G_{\mathit{e-l}} (T_e )\Delta T_e
\end{equation}
\begin{equation}
j\omega C_e \Delta T_e+G_{\mathit{e-l}}(T_e ) \Delta T_e %
= G_{\mathit{e-l}}(T_l )\Delta T_l - G_{\mathit{ETF}}\Delta T_e.
\label{e_ab2}
\end{equation}
If we want to build the block diagram associated with these three
equations we can consider the left side of the equations as the
response function of the three systems (absorber, lattice, electrons), 
and the right side as the input to each system.
Connecting the three systems gives the block diagram of Fig.~\ref{f_abs1}.
The diagram can be solved to obtain the detector responsivity:
\begin{equation}
\begin{array}{c}
S(\omega )= \frac{1}{j\omega C_a(1+j\omega \tau _e) + %
(1+j\omega \tau _e)(1+j\omega \tau _a)(G+G_{\mathit{e-l}}(T_l )+j\omega C_l) %
-G_{\mathit{e-l}} (T_e )A_{\mathit{e-l}} (1+j\omega \tau _a)}
\frac{A_{\mathit{e-l}} X \alpha A_{tr}}{T_e}
\end{array}
\end{equation}
with $\tau _a=C_a /G_a$. Notice that if
$G_a \rightarrow \infty$ this expressions reduces to
the one without absorber decoupling for a
detector with lattice heat capacity $C_l=C_l +C_a$.

Using Eqs.~\ref{e_alpha},~\ref{e_z1},~\ref{e_z2b}, and \ref{e_ab1} through
\ref{e_ab2}, we can also calculate the detector dynamic impedance:
\begin{equation}
\begin{array}{c}
Z(\omega )=R\frac{ %
[(G+G_{\mathit{e-l}}(T_l )+j\omega C_l)(1+j\omega \tau _a)+j\omega C_a] %
\left( G_{\mathit{e-l}}(T_e )+j\omega C_e +\frac{P\alpha}{T_e}\right) %
-G_{\mathit{e-l}}(T_e ) \cdot G_{\mathit{e-l}}(T_l )\cdot (1+j\omega \tau _a)
}{
[(G+G_{\mathit{e-l}}(T_l )+j\omega C_l)(1+j\omega \tau _a)+j\omega C_a] %
\left( G_{\mathit{e-l}}(T_e )+j\omega C_e -\frac{P\alpha}{T_e}\right) %
-G_{\mathit{e-l}}(T_e ) \cdot G_{\mathit{e-l}}(T_l )\cdot (1+j\omega \tau _a)
}
\end{array}
\end{equation}

\subsection{Noise contribution}

As in the hot electron model of the thermometer, there are two
effects introduced by the thermal link between the absorber
and the lattice system. The first effect is that the
response of the detector is different, therefore the
$NEPs$ due to thermal, Johnson, $1/f$, load resistor,
amplifier, and hot-electron noise are different. The second effect
is the introduction of an extra noise term due to the power
fluctuations between absorber and lattice.

\begin{figure}
\includegraphics{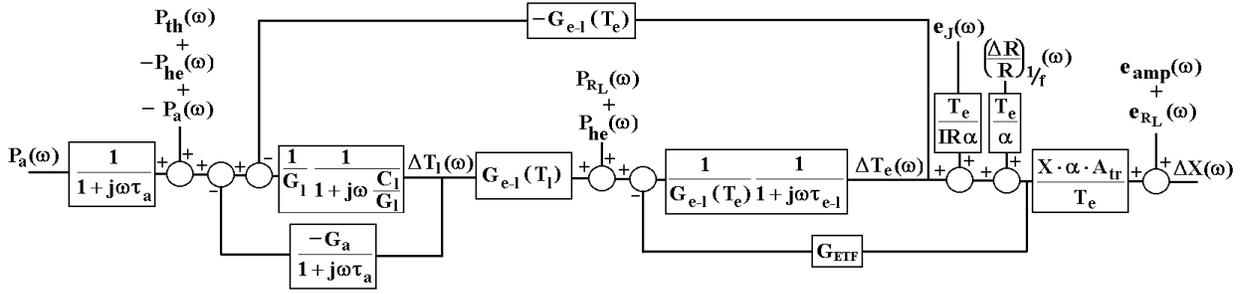}
\caption{\label{f_abs2}
Block diagram representation of noise in a detector
with a finite thermal conductivity between absorber
and lattice system.}
\end{figure}

Fig.~\ref{f_abs2} shows the block diagram of the
detector with the noise sources evident. As in the hot electron 
model, the noise
due to the link between absorber and lattice can be 
described as a power flow out of the absorber and into the 
lattice or viceversa. This power has same
value $P_a$, but opposite sign at the two ends of the link. 
Since the temperature of absorber and lattice systems are 
equal, the value of
$P_a$ is simply:
\begin{equation}
P_a =\sqrt{4k_b G_a T_l ^2}.
\end{equation}

Solving the block diagram of Fig.~\ref{f_abs2}
independently for each noise source we obtain:
\begin{equation}
NEP_{th}=P_{th}(1+j\omega \tau _a)
\label{e_absn1}
\end{equation}
\begin{equation}
\begin{array}{c}
NEP_J =\sqrt{\frac{4k_b T_e}{P\alpha ^2}} \frac{T_e ^{\beta _e +1}} %
{T_l ^{\beta _e}} \left[ (1+j\omega \tau _{\mathit{e-l}}) \left( j\omega C_a + G%
(1+j\omega \tau _a)(1+j\omega \tau _l) \right) + %
(1+j\omega \tau _a)  \frac{T_l ^{\beta _e }}{T_e ^{\beta _e}} %
j\omega C_e \right]
\end{array}
\end{equation}
\begin{equation}
\begin{array}{rcl}
NEP_{1/f} & = & \left(\frac{\Delta R}{R}\right) _{1/f}\frac{T_e}{\alpha}
\frac{T_e ^{\beta _e }} {T_l ^{\beta _e}} \\
& \times & \left[ (1+j\omega \tau _{\mathit{e-l}}) \left( j\omega C_a + G%
(1+j\omega \tau _a)(1+j\omega \tau _l) \right)
+(1+j\omega \tau _a)  \frac{T_l ^{\beta _e }}{T_e ^{\beta _e}} %
j\omega C_e \right]
\end{array}
\end{equation}
\begin{equation}
NEP_{R_L} =\frac{e_{R_L}}{S(\omega )}+\frac{2Ie_{R_L}}{G_{\mathit{e-l}}(T_l )} %
\left[ j\omega C_a +(1+j\omega \tau _a) (G+G_{\mathit{e-l}}(T_l ) + %
j\omega C_l) \right]
\end{equation}
\begin{equation}
NEP_{amp}=e_{amp}/S(\omega )
\end{equation}
\begin{equation}
NEP_{he}=\frac{P_{he}}{G_{\mathit{e-l}}(T_l )} %
\left[ j\omega C_a +G(1+j\omega \tau _a)(1+j\omega \tau _l) \right]
\end{equation}
\begin{equation}
NEP_a =P_a j\omega \tau_a
\label{e_absn2}
\end{equation}
Notice again that if $G_a \rightarrow \infty$ these expressions
are equal to the hot-electron expressions for a detector
with lattice heat capacity $C_l +C_a$ and the absorber $NEP$ is
equal to zero.

\section{Non-ohmic Behavior of the  Thermometer}

Another effect that may change the performance of a detector is the
non-ohmic behavior of the thermometer, i.e., the thermometer resistance
may not depend only on the thermometer electron temperature, but also on
the current (or voltage) that is used to readout the temperature
change: $R=R(T_e ,I)$ \cite{mather84}. 
This effect is particularly strong when TES thermometers are used \cite{mark}.
The responsivity of a detector with non-ohmic thermometer has already been 
calculated by J.~C.~Mather in 1984 \cite{mather84} and its
effect on TES microcalorimeters has
been studied in detail by M.~A.~Lindeman in 2000 \cite{mark}.
A non-ohmic thermometer can also be easily included in our
model. If the resistance of the thermometer depends on the 
readout signal, we can write:
\begin{equation}
dR=\frac{R}{T_e}\alpha _I dT_e +\frac{R}{I} \beta _I dI
\label{nonlinear1}
\end{equation}
or equivalently
\begin{equation}
dR=\frac{R}{T_e}\alpha _V dT_e +\frac{R}{V} \beta _V dV,
\label{nonlinear2}
\end{equation}
where:
\begin{equation}
\alpha _I=\frac{T_e}{R}\frac{\partial R}{\partial T_e}\Bigg\vert _I, \; %
\beta _I=\frac{I}{R}\frac{\partial R}{\partial I}\Bigg\vert _{T_e}, \; %
\alpha _V=\frac{T_e}{R}\frac{\partial R}{\partial T_e}\Bigg\vert _V, \; %
\mathrm{and} \; %
\beta _V=\frac{V}{R}\frac{\partial R}{\partial V}\Bigg\vert _{T_e}.
\end{equation}
Using Eq.~\ref{nonlinear1} or Eq.~\ref{nonlinear2} is equivalent,
and it is always possible to go from one notation to the other
using Ohm's laws:
\begin{equation}
\beta _V =\frac{\beta _I}{1+\beta _I } \; %
\mathrm{and} \; %
\alpha _V = \frac{\alpha _I}{1+\beta _I}.
\end{equation}

The only terms in our model that are affected by the non-ohmic
behavior are the electro-thermal feedback term $G_{\mathit{ETF}}$ and
the transducer responsivity $A_{tr}$. We can calculate them
assuming the bias circuit of Fig.~\ref{f_model1}.
\begin{equation}
P=I^2 R \Rightarrow \Delta P=2IR\Delta I+I^2\Delta R,
\end{equation}
\begin{equation}
I=\frac{V_{bias}}{R_L +R} \Rightarrow \Delta I=-\frac{I}{R_L +R}\Delta R,
\end{equation}
and
\begin{equation}
V=V_{bias}-IR_L \Rightarrow \Delta V=-R_L \Delta I.
\end{equation}
Using Eq.~\ref{nonlinear1} we obtain:
\begin{equation}
\Delta P=-\frac{P}{T_e}\frac{R-R_L}{R_L +R(1+\beta _I)} \alpha _I \Delta T_e,
\end{equation}
\begin{equation}
\frac{\Delta V}{V}=\alpha _I \frac{R_L}{R_L +R(1+\beta _I)} \frac{\Delta T_e}{T_e},
\end{equation}
and
\begin{equation}
\frac{\Delta I}{I}=-\alpha _I \frac{R}{R_L +R(1+\beta _I)} \frac{\Delta T_e}{T_e}.
\end{equation}

The model describing a non-ohmic thermometer is therefore identical to that
describing a linear one, with the substitution:
\begin{equation}
\alpha \rightarrow \alpha _I
\end{equation}
\begin{equation}
G_{\mathit{ETF}}=\frac{P}{T_e}\frac{R-R_L}{R_L +R(1+\beta _I)} \alpha _I,
\end{equation}
and
\begin{equation}
A_{tr}=\frac{R_L}{R_L +R(1+\beta _I)}
\end{equation}
for voltage readout, or
\begin{equation}
A_{tr}=-\frac{R}{R_L +R(1+\beta _I)}
\end{equation}
for current readout.
With this substitution in the equations that we derived previously in the paper,
it is possible to predict both responsivity and noise in the detector. 

We can also use Eq.~\ref{nonlinear1} to calculate the dynamic impedance 
of the detector. In the case of absorber and hot-electron
decoupling we obtain:
\begin{equation}
\begin{array}{rcl}
Z(\omega ) & = & R(\beta_I +1) \\
& \times & \frac{ %
[(G+G_{\mathit{e-l}}(T_l )+j\omega C_l)(1+j\omega \tau _a)+j\omega C_a] %
\left( G_{\mathit{e-l}}(T_e )+j\omega C_e +\frac{P\alpha _I}{T_e (\beta _I +1)}\right) %
-G_{\mathit{e-l}}(T_e ) \cdot G_{\mathit{e-l}}(T_l )\cdot (1+j\omega \tau _a)
}{
[(G+G_{\mathit{e-l}}(T_l )+j\omega C_l)(1+j\omega \tau _a)+j\omega C_a] %
\left( G_{\mathit{e-l}}(T_e )+j\omega C_e -\frac{P\alpha _I}{T_e}\right) %
-G_{\mathit{e-l}}(T_e ) \cdot G_{\mathit{e-l}}(T_l )\cdot (1+j\omega \tau _a)
}.
\end{array}
\end{equation}
This reduces to:
\begin{equation}
\begin{array}{c}
Z(\omega )=R(\beta_I +1)\frac{ %
(G+G_{\mathit{e-l}}(T_l )+j\omega C_l) %
\left( G_{\mathit{e-l}}(T_e )+j\omega C_e +\frac{P\alpha _I}{T_e (\beta_I +1)}\right) %
-G_{\mathit{e-l}}(T_e ) \cdot G_{\mathit{e-l}}(T_l )
}{
(G+G_{\mathit{e-l}}(T_l )+j\omega C_l) %
\left( G_{\mathit{e-l}}(T_e )+j\omega C_e -\frac{P\alpha _I}{T_e}\right) %
-G_{\mathit{e-l}}(T_e ) \cdot G_{\mathit{e-l}}(T_l )
}
\end{array}
\end{equation}
for hot-electron decoupling only, and to:
\begin{equation}
Z(\omega )=R(\beta_I +1)\frac{1+\frac{P\alpha _I}{GT(\beta_I +1)}+j\omega \tau}
{1-\frac{P\alpha _I}{GT}+j\omega \tau}
\end{equation}
for the ideal model.

We do not know of a rigorous general method for deriving the 
Johnson noise in a non-ohmic resistor.  Nor does there seem to be a 
single definite scheme for determining the net response of the detector to 
this fundamental thermal noise, since it is an internal noise 
generated in the non-ohmic resistor, and it is not clear how it 
should itself affect the non-ohmicity of the resistor.  We are investigating this 
further, but for the present have assumed that the Johnson noise can 
be represented as a random voltage source with power spectral density 
$4k_b T_e R$ in series with the non-ohmic resistance, and that the 
Johnson fluctuations in the source cause the resistance to fluctuate due to 
the current dependence of the resistor.  This results in the same suppression 
of the Johnson noise due to the current dependence of the resistance as 
occurs for external signals 
and noise if the non-ohmic resistance is expressed as $R(T_e ,I)$. 
This uncertainty (or dependence on the details of the physics) 
applies only to the Johnson noise of the sensor.  Small-signal 
responsivities to all external sources of signal and noise are 
unambiguous, so it is only the detector Johnson noise contribution to 
the $NEP$ that is uncertain.

\section{Results}

To verify our results we simulated the performance of an existing 
microcalorimeter and compared the results with data from the detector. 
We considered a microcalorimeter used in the development phase
of the X-Ray Spectrometer (XRS) for the Astro-E satellite \cite{xrs}. 
The detector that we used for the comparison was part of a 
6x6 test array of microcalorimeters with silicon implanted
thermistors and HgTe absorbers. We chose this detector because
the array has been studied in great detail and the characteristics 
of the pixels are well known.

\begin{figure}
\includegraphics{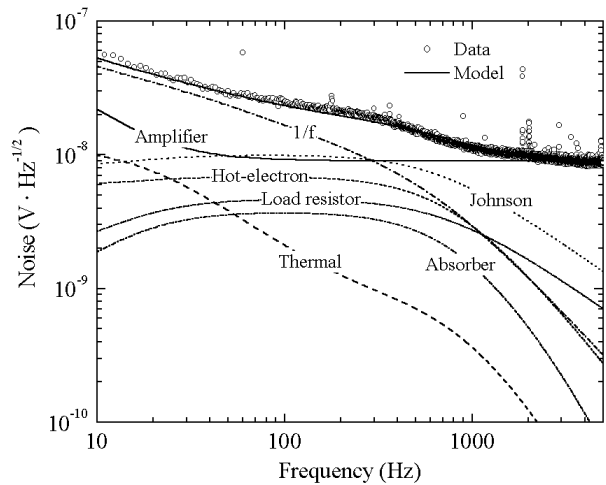}
\caption{\label{f_XRS}
Comparison between the noise from a 6x6 XRS array pixel 
(courtesy of Caroline K. Stahle) and our
model. The model includes the effect of the decoupling between hot
electrons and phonons in the sensor and between absorber and sensor.
The noise sources that are included are
Johnson noise of the sensor, thermal noise due to the link between 
detector and heat sink, thermal noise due to the link between phonons 
and electrons in the sensor, thermal noise due to the link between
absorber and sensor, Johnson noise of the load resistor, $1/f$ noise, 
and noise of the readout electronics.}
\end{figure}

We first used Eq.~\ref{e_hot1}~and~\ref{e_hot1b}
to calculate the expected equilibrium temperature of the detector. 
We then used Eqs.~\ref{e_absn1} through \ref{e_absn2} to calculated 
the expected noise spectra. The sum of these can be compared with 
the measured noise spectrum, as shown in Fig.~\ref{f_XRS}. 
In the model all the input parameters are fixed to the values measured 
experimentally. The only value that was
not available and that has been adjusted during the calculation of the
theoretical noise is the stray capacitance between gate and source 
of the FET electronics. The value of 5~pF obtained for the stray capacitance
is in good agreement with typical values for the FET amplifiers used in 
the measurement.
The agreement between the model and the measurement is very good.

The data set has been acquired at a heat sink temperature of 65~mK.
The model predicts an equilibrium temperature of 77~mK and, through
Eq.~\ref{e_nep1}, an energy resolution of 8.4~eV, to be compared with 
the measured values of 78~mK and 8.65~eV. 
The agreement is well within the accuracy
of the input parameters in the model and demonstrates the power of 
the model in predicting detector performance.

Our analytical model has also been compared with a model that uses matrix
notation to numerically solve the linearized differential equations of the
microcalorimeter \cite{tali}. The numerical model was developed
independently at the NASA/Goddard Space Flight Center to predict the performance
of more complex detectors \cite{tali}. Using the same parameter values,
the agreement between the two is within the numerical error in the 
implementation of the models \cite{maxltd9}.

\section{Conclusions}

We have developed an analytical model that predicts the behavior of 
microcalorimeters and bolometers. 
The model includes the effect of hot electrons in the
detector sensor, a thermal decoupling between absorber and sensor,
and the effect of a non-ohmic thermometer.
The model analytically predicts the detector responsivity and expected noise
under these conditions. 
The noise sources that are included in the model are the
Johnson noise of the sensor, the thermal noise due to the link between 
detector and heat sink, the thermal noise due to the link between phonons 
and electrons in the sensor, the thermal noise due to the link between
absorber and sensor, the Johnson noise of the load resistor, the $1/f$ noise
as thermal noise, and the noise of the readout electronics.
A comparison between the predictions of our model and data from 
a detector developed for the XRS instrument shows good agreement.

We also described a different way to analyze the performance of 
bolometers and microcalorimeters, using block diagram algebra. The 
formalism that we introduced can be applied to the description
of different detector configurations.

\section*{Acknowledgments}

We would like to thank Caroline Stahle and Enectali Figueroa-Feliciano
for the useful discussion in the development of the model and for supplying
the XRS data and the comparison with the numerical model.
We also would like to thank the other members of the microcalorimeter
groups at the University of Wisconsin and the the NASA/Goddard Space 
Flight Center for the useful discussion and suggestions.
We are grateful to the referee for a careful reading 
and important suggestions.

This work was supported by NASA grant NAG5-629.

\end{document}